\documentclass[useAMS,usenatbib]{mn2e}

\usepackage[active]{srcltx}
\usepackage{graphicx}
\usepackage{txfonts}
\usepackage{natbib}
\usepackage{multirow}
\usepackage{longtable}

\title[Characterisation of the magnetic field of the Herbig Be star HD 200775]{Characterisation of the magnetic field of the \\Herbig Be star HD 200775\thanks{Based on observations obtained at the Canada-France-Hawaii Telescope (CFHT) which is operated by the National Research Council of Canada, the Institut National des Sciences de l'Univers of the Centre National de la Recherche Scientifique of France, and the University of Hawaii}}
\author[E. Alecian et al.]
{E.~Alecian$^1$$^2$\thanks{E-mail: evelyne.alecian@rmc.ca},
 C.~Catala$^2$,
 G.A.~Wade$^1$,
 J.-F.~Donati$^3$,
 P. Petit$^3$,
 J.D.~Landstreet$^4$,
 \newauthor
 S.~Bagnulo$^6$,
 T.~B\"ohm$^3$,
 J.-C.~Bouret$^5$,
 C.~Folsom$^6$ ,
 J.~Grunhut $^1$ and
 J.~Silvester$^1$ \\
 $^1$Dept. of Physics, Royal Military College of Canada, PO Box 17000, Stn Forces, Kingston K7K 7B4, Canada \\
 $^2$Observatoire de Paris, LESIA, 5, place Jules Janssen, F-92195 Meudon Principal CEDEX, France \\
 $^3$Laboratoire d'Astrophysique, Observatoire Midi-Pyr\'en\'ees, 14 avenue Edouard Belin, F-31400 Toulouse, France \\
 $^4$Dept. of Physics \& Astronomy, University of Western Ontario, London N6A 3K7, Canada \\
 $^5$Laboratoire d'Astrophysique de Marseille, Traverse du Siphon, BP8-13376 Marseille Cedex 12, France \\
 $^6$Armagh Observatory, College Hill, Armagh BT61 9DG, Northern Ireland, UK
}

\begin{document}

\date{Accepted . Received ; in original form }

\pagerange{\pageref{firstpage}--\pageref{lastpage}} \pubyear{2002}

\maketitle

\label{firstpage}

\begin{abstract}
The origin of the magnetic fields observed in some intermediate mass and high mass main sequence stars is still a matter of vigorous debate. The favoured hypothesis is a fossil field origin, in which the observed fields are the condensed remnants of magnetic fields present in the original molecular cloud from which the stars formed. According to this theory a few percent of the PMS Herbig Ae/Be star should be magnetic with a magnetic topology similar to that of main sequence intermediate-mass stars. 

After our recent discovery of four magnetic Herbig stars, we have decided to study in detail one of them, HD 200775, to determine if its magnetic topology is similar to that of the main sequence magnetic stars. With this aim, we monitored this star in Stokes $I$ and $V$ over more than two years, using the new spectropolarimeters ESPaDOnS at CFHT, and Narval at TBL. 

By analysing the intensity spectrum we find that HD 200775 is a double-lined spectroscopic binary system, whose secondary seems similar, in temperature, to the primary. We have carefully compared the observed spectrum to a synthetic one, and we found no evidence of abundance anomalies in its spectrum. We infer the luminosity ratio of the components from the Stokes $I$ profiles. Then, using the temperature and luminosity of HD 200775 found in the literature, we estimate the age, the mass and the radius of both components from their HR diagram positions. From our measurements of the radial velocities of both stars we determine the ephemeris and the orbital parameters of the system.

A Stokes $V$ Zeeman signature is clearly visible in most of the Least Square Deconvolution profiles and varies on a timescale on the order of one day. We have fitted the 30 profiles simultaneously, using a $\chi^2$ minimisation method, with a centered and a decentered-dipole model. The best-fit model is obtained with a reduced $\chi^2=1.0$ and provides a rotation period of $4.3281 \pm 0.0010$~d, an inclination angle of $60 \pm 11^{\circ}$, and a magnetic obliquity angle $\beta=125 \pm 8^{\circ}$. The polar strength of the magnetic dipole field is $1000 \pm 150$~G, which is decentered by $0.05 \pm 0.04$~$R_*$ from the center of the star. The derived magnetic field model is qualitatively identical to those commonly observed in the Ap/Bp stars.

Our determination of the inclination of the rotation axis leads to a radius of the primary which is smaller than that derived from the HR diagram position. This can be explained by a larger intrinsic luminosity of the secondary relative to the primary, due to a larger circumstellar extinction of the secondary relative to the primary.
\end{abstract}

\begin{keywords}
Stars : magnetic field -- Stars: pre-main-sequence -- Stars: binaries: spectroscopic -- Star: individual: HD 200775 -- Instrumentation : spectropolarimetry.
\end{keywords}

\begin{table*}
\caption{Log of the observations. Columns 1 and 2 give the UT date and the Heliocentric Julian Date of the observations. Column 3 gives the total exposure time. Column 4 gives the peak signal to noise ratio (at $\sim730$~nm per ccd pixel) in the spectra and column 5 gives the signal to noise ratio in the LSD Stokes $V$ profiles. Column 6 gives the longitudinal magnetic field, column 7 gives the rotation phase derived in Section 6.2, and columns 8 and 9 give the radial velocities of both components of the system. Column 10 gives the instrument used.}
\label{log}
\centering
\begin{tabular}{lccccr@{$\pm$}lcr@{$\pm$}lr@{$\pm$}ll}
\hline
Date   & HJD           & $t_{\rm exp}$ & S/N & S/N  & \multicolumn{2}{c}{$B_{\ell}$} & phase & \multicolumn{2}{c}{${\rm v}_{\rm radA}$} & \multicolumn{2}{c}{${\rm v}_{\rm radB}$} & Instrument\\
UT Time & (2 450 000+) & (s)     &       & (LSD)  & \multicolumn{2}{c}{(G)}    & & \multicolumn{2}{c}{(km.s$^{-1}$)} & \multicolumn{2}{c}{(km.s$^{-1}$)} & \\
\hline
 24/09/04 10:33 & 3272.9398 & 1200 & 480 & 1810 & -299 &   89 & 0.90 &    8.2   & 0.5   & -21.1 & 1.0 & ESPaDOnS \\
 22/05/05 14:35 & 3513.1072 & 3600 & 540 & 1940 &    91 &   84 & 0.39 & -19.3   & 1.1   &    1.2 & 3.6 & ESPaDOnS \\
 23/05/05 12:22 & 3514.0150 & 3600 & 420 & 1470 &    73 & 104 & 0.60 & -19.8   & 1.0   &    2.9 & 3.5 & ESPaDOnS \\
 24/05/05 12:09 & 3515.0065 & 3600 & 560 & 2090 & -262 &   78 & 0.83 & -19.9   & 1.0   &    0.0 & 3.4 & ESPaDOnS \\
 24/05/05 13:11 & 3515.0490 & 2400 & 470 & 1690 & -376 &   98 & 0.84 & -20.0   & 1.0   &    0.8 & 3.4 & ESPaDOnS \\
 25/05/05 12:15 & 3516.0101 & 3600 & 520 & 1880 & -411 &   78 & 0.06 & -19.9   & 1.0   &    0.9 & 3.3 & ESPaDOnS \\
 25/05/05 14:02 & 3516.0511 & 2400 & 300 & 1660 & -306 &   88 & 0.07 & -20.3   & 1.0   &   -0.9 & 3.4 & ESPaDOnS \\
 20/06/05 13:23 & 3542.0596 & 800   & 260 & 840   & -583 & 180 & 0.08 & -21.3   & 0.9   &    2.3 & 3.1 & ESPaDOnS \\
 22/06/05 13:43 & 3544.0732 & 800   & 240 & 540   &  229 & 199 & 0.54 & -21.5   & 0.9   &    1.5 & 3.0 & ESPaDOnS \\
 25/06/05 13:42 & 3547.0730 & 800   & 280 & 920   & -378 & 188 & 0.24 & -21.2   & 1.0   &    3.0 & 3.4 & ESPaDOnS \\
 19/07/05 11:43 & 3570.9915 & 1200 & 370 & 1330 &    37 & 118 & 0.76 & -22.6   & 0.9   &    4.7 & 3.2 & ESPaDOnS \\
 20/07/05 10:23 & 3571.9358 & 1200 & 530 & 1830 & -293 &   90 & 0.98 & -21.6   & 1.9   &    2.9 & 6.8 & ESPaDOnS \\
 26/08/05 10:00 & 3608.9206 & 1600 & 610 & 2280 &  156 &   73 & 0.53 & -23.3   & 1.0   &    5.0 & 3.4 & ESPaDOnS \\
 09/06/06 10:39 & 3895.9447 & 1200 & 550 & 1970 & -404 &   82 & 0.84 & -20.2   & 1.0   &    1.9 & 3.2 & ESPaDOnS \\
 10/06/06 14:31 & 3897.1059 & 2160 & 780 & 2840 & -383 &   57 & 0.11 & -19.8   & 1.0   &    0.0 & 3.2 & ESPaDOnS \\
 11/06/06 14:03 & 3898.0867 & 2160 & 710 & 2590 &   -24 &   69 & 0.33 & -20.4   & 1.0   &    4.4 & 3.2 & ESPaDOnS \\
 12/06/06 10:30 & 3898.9386 & 2160 & 760 & 2720 &  161 &   63 & 0.53 & -20.2   & 0.9   &    3.2 & 3.2 & ESPaDOnS \\
 13/06/06 10:08 & 3899.9231 & 2400 & 790 & 2780 & -193 &   60 & 0.76 & -20.1   & 0.9   &    1.1 & 2.9 & ESPaDOnS \\
 13/06/06 10:52 & 3899.9538 & 2400 & 810 & 2930 &   -97 &   59 & 0.77 & -20.2   & 0.8   &    0.7 & 2.8 & ESPaDOnS \\
 14/06/06 15:18 & 3901.1387 & 800   & 540 & 2000 & -269 &   79 & 0.04 & -20.4   & 1.1   &    2.6 & 3.8 & ESPaDOnS \\
 15/06/06 15:13 & 3902.1352 & 1840 & 760 & 2820 & -146 &   58 & 0.27 & -19.9   & 1.1   &    8.6 & 4.1 & ESPaDOnS \\
 16/06/06 15:14 & 3903.1365 & 1720 & 740 & 2700 &  180 &   64 & 0.50 & -21.5   & 1.3   &    9.3 & 4.7 & ESPaDOnS \\
 10/12/06 4:27   & 4079.6850 & 1600 & 330 & 1150 &   -93 & 132 & 0.29 & -14.2   & 0.9   &   -0.5 & 3.3 & ESPaDOnS \\
 10/12/06 6:16 & 4079.7610 & 1600   & 220 & 770   & -198 & 188 & 0.31 & -14.0   & 1.0   &   -2.7 & 3.5 & ESPaDOnS \\
 11/12/06 5:09 & 4080.7141 & 1200   & 470 & 1860 &  320 &   83 & 0.53 & -13.7   & 0.9   &   -2.9 & 3.4 & ESPaDOnS \\
 11/12/06 5:32 & 4080.7302 & 1200   & 470 & 1800 &    34 &   85 & 0.53 & -13.9   & 0.9   &   -3.4 & 3.2 & ESPaDOnS \\
 04/03/07 16:00 & 4164.1633 & 1320 & 600 & 2130 & -198 &   58 & 0.81 &   -9.9   & 0.8   &   -6.1 & 3.0 & ESPaDOnS \\
 24/04/07 4:09   & 4214.6708 & 2700 & 410 & 1460 &    17 &   95 & 0.48 &   -7.3   & 2.1   & -10.2 & 8.2 & Narval         \\
 26/04/07 4:09   & 4216.6713 & 2500 & 180 & 660   & -341 & 180 & 0.94 &   -7.5   & 2.7   &   -9.3 & 10.0 & Narval         \\
 26/06/07 12:58 & 4278.0424 & 2400 & 620 & 2280 & -233 &   52 & 0.12 &   -1.8   & 1.0   &   -9.1 & 3.4 & ESPaDOnS \\
\hline
\end{tabular}
\end{table*}

\section{Introduction}

Some main sequence A, B and O stars host strong ($\sim$ kG) organised magnetic fields. The origin of the magnetic fields of these intermediate and high mass stars is still a matter of debate. The favoured theory is the fossil field hypothesis. This theory assumes that the magnetic fields observed in these main sequence stars are relics of the magnetic fields which existed in the molecular clouds from which the stars formed. This theory implies that the remnant interstellar magnetic field should subsist throughout all the processes of formation encountered by the star, from the gravitational collapse to the main sequence phase, without being regenerated. According to this theory, some pre-main sequence (PMS) stars of intermediate mass should host magnetic fields.

For a long time, this theory conflicted with the general belief that all stars pass through a completely convective phase during the PMS phase. The turbulent diffusion produced by the convection would dissipate the magnetic field during the Hayashi phase. However \citet{palla93} calculated the birthline (the locus in the HR diagram where stars become optically visible and start the PMS phase), showing that the Hayashi phase is considerably reduced for stars between 1.5 $M_{\odot}$ and 2$M_{\odot}$ and disappears completely for stars above 2$M_{\odot}$. A fundamental conclusion of this work was that the magnetic field of the intermediate PMS stars can potentially survive the PMS phase.

According to the fossil field hypothesis, some Herbig Ae/Be stars (PMS stars of intermediate mass) should be magnetic. Many authors have tried to detect a magnetic field in these stars. \citet{catala93} observed AB Aur using a Zeeman polarimeter placed before the Coud\'e spectrograph of the CFHT. They found no circular polarisation signal in the FeII 5018 \AA $\;$ line and they obtained an upper limit around 1 kG. \citet{catala99} tried again to detect a magnetic field in AB Aur using the MUSICOS instrument temporarily installed at the CFHT. They observed no Zeeman signature and obtained a lower detection limit around 300 G. \citet{donati97} observed a large sample of cool and hot stars, including two Herbig stars, using the UCLES spectrograph of the AAT and the SemelPol polarimeter. Applying the Least Squares Deconvolution method, they detected a magnetic field of around 50 G in the Herbig Ae star HD 104237, but they obtained no detection in another Herbig Ae/Be star, HD 100546. \citet{hubrig04,hubrig07} claim detections in HD 139614  and HD 144432. Finally recent search for magnetic field in the field Herbig Ae/Be stars by \citet{wade07} has been carried out using the FORS1 spectropolarimeter at the ESO VLT. They identified two possible magnetic stars (HD~101412 and BF Ori), of which one has been confirmed (HD~101412), with higher resolution data (Wade et al. 2007 in preparation).

Recently, the new generation instrument ESPaDOnS was installed at the Cassegrain focus of the CFHT (Donati et al., in preparation). Thanks to the high efficiency and qualities of this high resolution spectropolarimeter, magnetic fields in several Herbig stars have been discovered. During the technical run of ESPaDOnS a magnetic field was discovered in the Herbig Ae/Be star HD 200775 (Donati et al., in preparation). Then, during scientific nights of the first semester of ESPaDOnS, three other stars were detected as magnetic : HD 72106, V380 Ori and HD 190073 \citep{wade05,catala07}. These recent discoveries bring new arguments to support the fossil field theory. However further exploration is necessary.

According to the fossil field hypothesis, the structure and the intensity of the magnetic field of these PMS stars should be similar to those of the main sequence magnetic stars. In order to verify this, we have observed HD 200775, with ESPaDOnS, during many nights. Using the temporal variations of the Stokes $V$ profile we were able to model the structure and the geometry of the magnetic field. The next section presents the observations and data reduction procedures. In Sect. 3, 4 and 5 we present the study of the intensity spectrum and in Section 6 the study of the polarization spectrum. Conclusions are given in Sect. 7.

%

\section{Observations and data reduction}

Our data were obtained using the high resolution spectropolarimeter ESPaDOnS installed on the 3.6 m Canada-France-Hawaii Telescope (Donati et al., in preparation) during many scientific runs. Table \ref{log} presents the log of the observations. One of our spectra has also been obtained using the instrument Narval, which is a copy of ESPaDOnS, installed on the 2 m Bernard Lyot Telescope (TBL) at the Pic du Midi observatory in France.

We used the instruments ESPaDOnS and Narval in polarimetric mode, and we obtained spectra with resolving power of 65000. Each exposure was divided in 4 sub-exposures of equal time in order to compute the optimal extraction of the polarisation spectra (Donati et al. 1997, Donati et al., in preparation). We recorded only circular polarisation, as the Zeeman signature expected in linear polarisation is about one order of magnitude lower than circular polarisation. The data were reduced using the "Libre ESpRIT" package especially developed for ESPaDOnS and Narval, and installed at the CFHT and at the TBL (Donati et al. 1997, Donati et al., in preparation). After reduction, we obtained the intensity Stokes $I$ and the circular polarisation Stokes $V$ spectra of the star observed, both normalised to the continuum intensity of HD 200775. A null spectrum ($N$) is also computed in order to diagnose spurious polarisation signatures, and to help to verify that the signatures in the Stokes $V$ spectrum are real \citep{donati97}.

The data of May 2005 were affected by a 1.3 mag loss compared to the data obtained more recently with ESPaDOnS. This problem, which was due to damage to the external jacket of the optical fibres, was fixed prior to the July run. In order to compensate for this damage and to obtain a satisfactory S/N ratio, we exposed longer (see Table \ref{log}).

We then applied the Least Squares Deconvolution procedure to all spectra \citep{donati97}. This method assumes that all selected lines of the intensity spectrum have a profile of similar shape. Hence, this supposes that all lines are broadened in the same way. We can therefore consider that the observed spectrum is a convolution between a profile (which is the same for all lines) and a mask including all choosen lines of the spectrum. We therefore apply a deconvolution to the observed spectrum using the mask, in order to obtain the average photospheric profiles of Stokes $I$ and $V$. In this procedure, each line is weighted by its signal to noise ratio, its depth in the unbroadened model and its Land\'e factor. The mask was first computed using Kurucz ATLAS 9 models \citep{kurucz93} with $T_{\rm eff}= 19000$ K and $\log g=3.5$ suitable for the star (Table \ref{fp}). We excluded from this mask hydrogen Balmer lines, strong resonance lines and lines whose Land\'e factor is unknown. Then we cleaned the mask by selecting the lines with a depth less than 0.4, in order to eliminate the lines contaminated by strong emission. We have also modified the line depths in order to take into account the relative depth of the lines of the observed spectrum. The final mask contains only 37 lines, but is sufficent to compute the LSD Stokes $I$ and $V$ profiles with a good S/N ratio (Table \ref{log}). To explore the sensitivity of the LSD profiles to the detailed line list, line masks were constructed using weak lines including He lines, and using weak lines excluding He lines. The various LSD profiles were analysed in the manner described in Section 3. No significant differences were found. The null $N$ profile has been computed in the same way, from the null spectrum, and will be used, in Sect. 6, to check that the signature in the Stokes $V$ profile is real.

The LSD average line profiles were computed on a velocity grid with 3.6 km.s$^{\rm -1}$ sampling. The resulting relative noise in the LSD Stokes $V$ profiles is given in the 5th column of Table \ref{log}.


\section{Binarity}

\begin{figure*}
\centering
\includegraphics[width=9cm,angle=90]{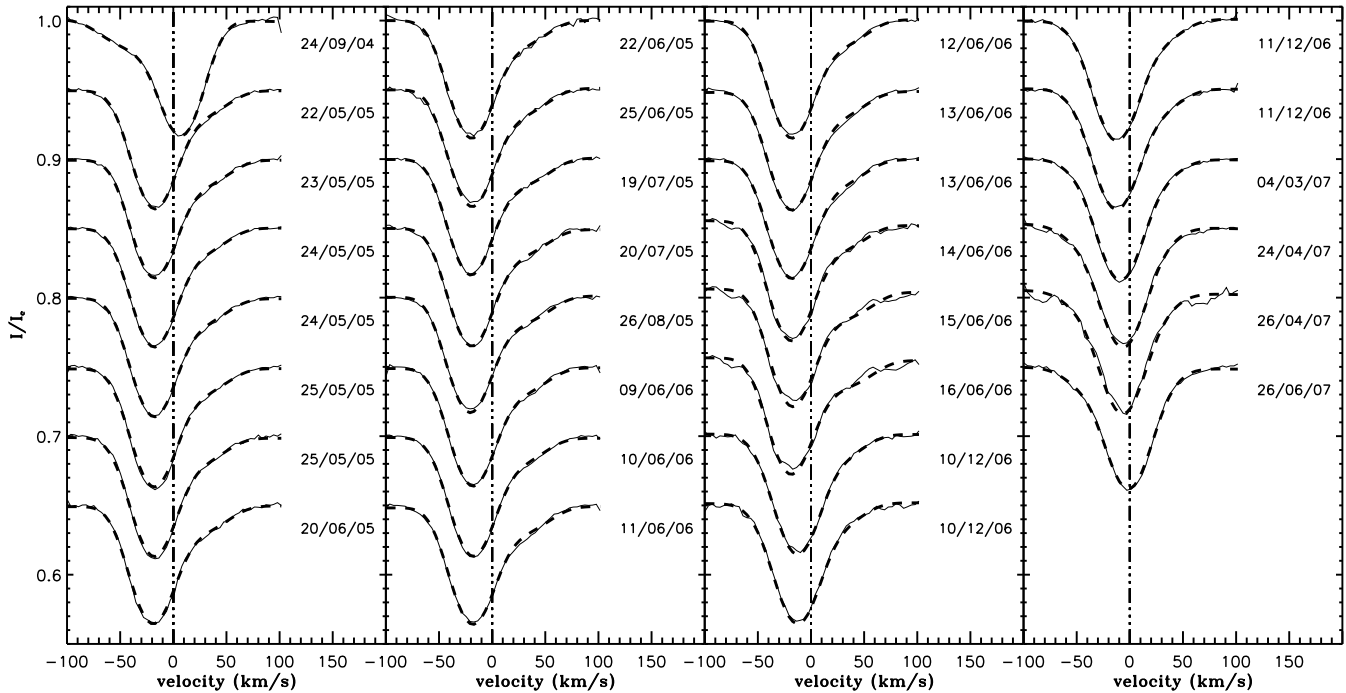}
\caption{LSD Stokes $I$ profiles of all spectra (thin full line) superimposed with their fit (thick dashed-line) described in Section 3.}
\label{allI}
\end{figure*}

In the litterature, many authors refer to the binarity of HD~200775. \citet{corporon99} detected radial velocity variations of photospheric lines of the spectrum of HD 200775, but they were not able to determine the orbital period. \citet{miroshnichenko98} analysed spectroscopic data of this star obtained by many authors over 20 years. They noted a cyclic variation of the equivalent width of H$\alpha$ with a 1345 day period, and they suggested that a companion star could be the trigger of the H$\alpha$ activity. \citet{pogodin04} discussed again this theory using new data and observational evidence of radial velocity variations of some photospheric lines. They plotted two radial velocity curves, obtained on the one hand from the H$\alpha$ wings, and on the other hand from photospheric lines. They fitted them with a synthetic radial velocity curve in order to obtain the ephemeris and the orbital parameters of the double system. Finally \citet{monnier06} were able to separate both stars of HD~200775 using interferometry, and determine orbital parameters for the system.

\begin{figure}
\centering
\resizebox{\hsize}{!}{\includegraphics[clip=true]{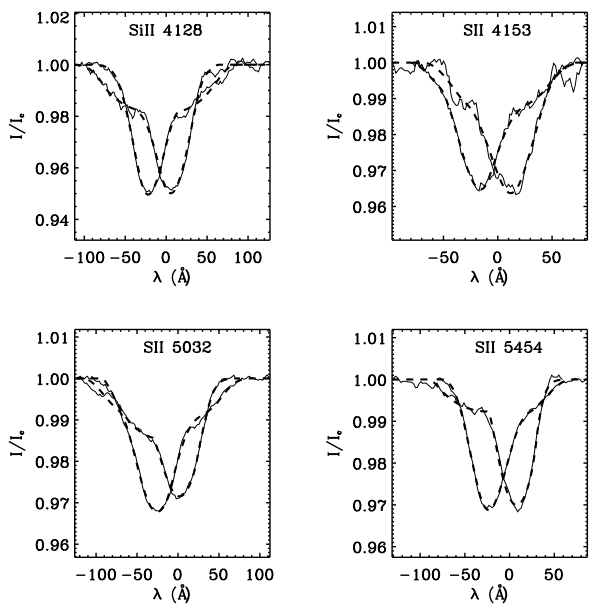}}
\caption{Photospheric lines of SiII 4128\AA, SII 4153\AA, SII 5032\AA, and SII 5454\AA, oberved on September 23rd 2004 (redshifted) and on August 25th 2005 (blueshifted). The fit (thick dashed line) with a convolution of a gaussian and a rotation profile is superimposed to the observed profiles (thin full line).}
\label{phot}
\end{figure}

Figure \ref{allI} shows the LSD average Stokes $I$ profiles (thin full line) of all our observations, obtained using the cleaned mask as explained in Section 2. We can see the presence of a second component in each profile : on the blue for the Sept. 2004 profile and on the red for the others. Furthermore, Fig. \ref{phot} shows some individual absorption lines of the intensity spectrum, observed in Sept. 2004 and May 2005, where the secondary component is visible. Between Sept. 2004 and May 2005 the primary component of the profile and the photospheric lines shifted towards the blue, while the secondary component moved in the other direction. We therefore assign this second component of the average profile to a companion star. HD 200775 is therefore an SB2 system whose orbital period is likely greater than one year. In the following we call HD 200775A (with the shaper, deeper lines) the primary component and HD 200775B (with the broader, shallower lines) the secondary component. 

In order to study the variations of the radial velocities of our spectra, we performed a simultaneous Least-Squares fit to the 30 LSD $I$ profiles\footnote{Although a potentially more effective approach would be to perform spectral disentangling, our data set was not suitable for such an analysis due to the sparse coverage of the orbital period and the small velocity separation of the lines of the components  (O. Kochukhov, private communication).}. Each profile is fitted with the sum of two functions, each function modeling the line profile of one components. Each one of these two functions is the convolution of a rotation function (for which the projected rotational velocity ${\rm v}\sin i$ is a free parameter in the fit) and a gaussian whose width is fixed and computed from the spectral resolution and the inferred macroturbulent velocity \citep{gray92}. We adopted an isotropic macroturbulent velocity of 15 km.s$^{-1}$ in order to fit the wings of the profiles of individual lines in the spectrum (both strong and weak lines), and of the LSD $I$ profiles.

\begin{figure}
\centering
\includegraphics[width=8cm]{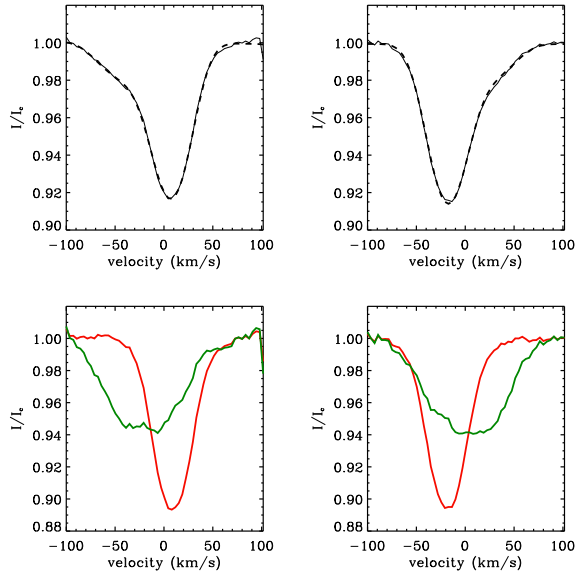}
\caption{{\it Upper panels:} LSD Stokes $I$ profiles from spectra obtained on September 23, 2004 and September 21, 2005 (full line). The fitted line profiles are the superimposed dashed lines. {\it Lower panels:} Individual intensity profiles of the primary (narrow red profile) and secondary (large green profile) components of the system.}
\label{rotprof}
\end{figure}

The free parameters of the fitting procedure are the centroids, depths, and projected rotational velocities (${\rm v}\sin i$) of both components. The centroids of both functions can vary from one profile to another., whereas the depths and ${\rm v}\sin i$ of both components cannot. This fitting procedure therefore assumes that the depths and ${\rm v}\sin i$ of both components do not vary with time, which we confirm (within the error bars) by fitting each profile separately. Figure \ref{rotprof} shows an example of such a fit for the Sept. 2004 and 22 May 2005 profiles, as well as the profiles of the isolated components calculated by subtracting the synthetic profile of the other component from the observed LSD Stokes $I$ profile. This automatic fitting procedure enables us to measure the ${\rm v}\sin i$ and radial velocities of both components. We obtain projected rotational velocities of $26\pm2$ km.s$^{\rm -1}$ and $59\pm5$ km.s$^{\rm -1}$ for the primary and the secondary components, respectively. Assuming an orbital period greater than one year (consistent with our data) and considering a projected rotational velocity of the primary of 26 km.s$^{\rm -1}$, we conclude that the system is not synchronised.

The inferred macroturbulent velocity is not determined very precisely: acceptable values range from about 10-30~km.s$^{\rm -1}$. However, the fitting procedure is largely insensitive to these uncertainties - changing the macroturbulence within the uncertainties results in only very small changes to the inferred ${\rm v}\sin i$s, and to no significant change in the radial velocities.

\begin{figure}
\centering
\includegraphics[width=8cm,angle=90]{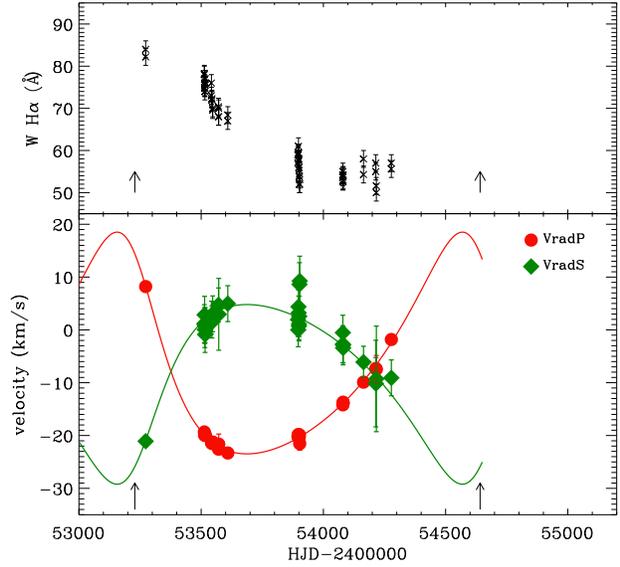}
\caption{{\it Upper panel} : Equivalent width of H$\alpha$ in function of time. H$\alpha$ appears in two different orders in the spectrum. Therefore the values measured on both orders are plotted. {\it Lower panel} : Radial velocities of the primary (red filled circle) and the secondary (green filled diamond) components of HD 200775. The full lines are the fitting curves. The arrows indicate the periastron passage.}
\label{fitvrad}
\end{figure}

We observe a strong variation of the radial velocity of the photospheric lines of the primary star between Sept. 2004 and May 2005 from $\sim+7$~km.s$^{\rm -1}$ to $\sim-20$ km.s$^{\rm -1}$. Then it is slightly varying, and since August 2005 it has been increasing slowly. In Fig. \ref{fitvrad} (lower panel), we plot the radial velocity of both components as a function of time, fitted with the radial velocity curves of an eccentric binary system, using a $\chi^2$ minimization method. We obtain the following parameters for the system: orbital period $P = 1412 \pm 54$ d, periastron epoch $T_0 = 2448991 \pm 152$ d, systemic radial velocity $\gamma = -7.9 \pm 0.9$ km.s$^{-1}$, eccentricity $e = 0.32 \pm 0.06$, periastron longitude $\omega = 216 \pm 12^{\circ}$, and radial velocity amplitude of both components $K_{\rm P}=20.9 \pm 2.5$ km.s$^{-1}$ and $K_{\rm S}=17.0 \pm 2.5$ km.s$^{-1}$, leading to the mass ratio of the system $q=\frac{M_{\rm P}}{M_{\rm S}}=\frac{K_{\rm S}}{K_{\rm P}}=0.81 \pm 0.22$.

\begin{figure}
\centering
\includegraphics[width=5cm,angle=90]{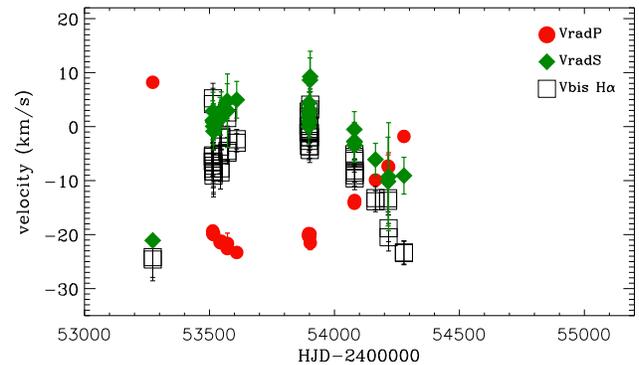}
\caption{Radial velocities of the primary (red filled circle) and the secondary components (green filled diamond) of HD 200775 superimposed with the bisector velocities of the $H\alpha$ line (black open square), measured at a level of $1.5-2.0 F_{\rm c}$. Note that the bisector velocities trace the radial velocities of the secondary, and clearly do not follow those of the primary.}
\label{plotvbis}
\end{figure}

\begin{table*}
\caption{Orbital parameters of the system found in this work, compared to previous work \citep{pogodin04,monnier06}.}
\label{tab:orb}
\centering
\begin{tabular}{llll}
\hline\hline
Parameter   & Radial Velocity & Bisector velocity & Interferometry  \\
                    & This work         & (Pogodin et al. 2004)  & (Monnier et al. 2006)\\
\hline
 & Results of RV fit & & \\
Period (days) & $1412 \pm 54$ & $1341 \pm 23$ & $1377 \pm 25$ \\
$T_0$ (HDJ) & $2448991 \pm 152$ & $2449149 \pm 87$ & $2449152 \pm 90$ \\
$\gamma$ (km/s) & $-7.9 \pm 0.9$ & $5.0 \pm 0.6$ & \\
e & $0.32 \pm 0.06$ & $0.29 \pm 0.07$ & $0.30 \pm 0.06$ \\
$\omega$ ($^{\circ}$) & $216 \pm 12$ & $203 \pm 22$ & $224 \pm16$ \\
$K_{\rm P}$ (km/s) & $20.9 \pm 2.5$ & & \\
$K_{\rm S}$ (km/s) & $17.0 \pm 2.5$ & $11.2 \pm 0.7$ & \\
 & & & \\
 & Derived parameters& & \\
$a\sin  i_{\rm orb}$ (AU) & $4.9 \pm 0.8$ & & \\
$a_{\rm P}\sin  i_{\rm orb}$ (AU) & $2.73 \pm 0.46$ & & \\
$a_{\rm S}\sin  i_{\rm orb}$ (AU) & $2.21 \pm 0.38$ & $1.38 \pm 0.11$ & \\
$M_{\rm P}\sin^3 i_{\rm orb}$ ($M_{\odot}$) & $3.6 \pm 1.6$ & & \\
$M_{\rm S}\sin^3 i_{\rm orb}$ ($M_{\odot}$) & $4.4 \pm 1.9$ & & \\
$i_{\rm orb}$ ($^{\circ}$) & $48^{+17}_{-13}$ & & $65 \pm 8$ \\
$a$ (AU) & $6.7 \pm 1.9$ & & \\
$a$ (mas) & $16 \pm 9$ & & $15.14 \pm 0.70$ \\
\hline
\end{tabular}
\end{table*}

The orbital parameters that we found are similar to those obtained by \citet{pogodin04} from the bisector velocities of H$\alpha$. In  order to compare our work to theirs, we also measured the bisector velocities of the H$\alpha$ line, as well as the other emission lines of the intensity spectrum. We observe that all of them closely follow the radial velocity of the secondary star (see Fig. \ref{plotvbis}). We therefore conclude that the region from which these emission lines originate is linked to the secondary component of the system. These results are in contradiction with those of Pogodin et al. who found that the bisector velocity follows the radial velocity of the primary star. Whether this contradiction is related to a real change of behavior of the binary system and its environment, or to problems of measurement of bisector and/or photospheric radial velocities in the data of Pogodin et al., remains to be determined.

Assuming that the variations of the bisector velocities of H$\alpha$ originate from the orbital movement of the secondary, we included them into the fit of the radial velocities. We found no significant differences of the inferred orbital parameters compared to those obtained from the fit of solely the radial velocities.

Using our determination of the orbital parameters we can derive the semi-major axis of the orbit, and the masses of the primary and the secondary, as functions of the inclination of the orbit: $a\sin i_{\rm orb}$, $M_{\rm P}\sin^3i_{\rm orb}$, and $M_{\rm S}\sin^3i_{\rm orb}$ (see Table \ref{tab:orb}). Using our determination of $M_{\rm P}$ and $M_{\rm S}$ (Table \ref{fp}) from stellar evolutionary models, we can therefore estimate the orbital inclination of the system. We find $\sin i_{\rm orb}{\rm(P)} = 0.70 \pm 0.16$ from $M_{\rm P}$ and $M_{\rm P}\sin^3i_{\rm orb}$, and $\sin i_{\rm orb}{\rm(S)}=0.78 \pm 0.17$ from $M_{\rm S}$ and $M_{\rm S}\sin^3i_{\rm orb}$. We determine the final value of $\sin i_{\rm orb}$ by taking a weighted avarage of $\sin i_{\rm orb}{\rm (P)}$ and $\sin i_{\rm orb}{\rm (S)}$. We obtain $\sin i_{\rm orb} = 0.74 \pm 0.17$, leading to $i_{\rm orb} = 48^{+17}_{-13}$~$^{\circ}$. Then we use this value to derive the semi-major axis of the orbit : $a = 6.7 \pm 1.9$~AU. Using the Hipparcos parallax of HD~200775 ($\pi = 2.33 \pm 0.62$~mas),  this corresponds to a projected separation of the components $a = 16 \pm 9$~mas.

The values of the orbital parameters that we derived are summarised in Table \ref{tab:orb} and compared to the orbital elements of Pogodin et al. (2004), as well as to the orbital parameters determined from interferometric data by \citet{monnier06}. Most of our measurements of the orbital parameters agree well with both works. Our determination of $K_{\rm S}$ is larger than that of Pogodin et al. which is likely due to our lack of data during the minimum separation of both stars. We need to continue to observe the system in order to confirm that value.


\section{Fundamental parameters of the primary and the secondary}

\begin{figure*}
\centering
\includegraphics[width=10cm,angle=90]{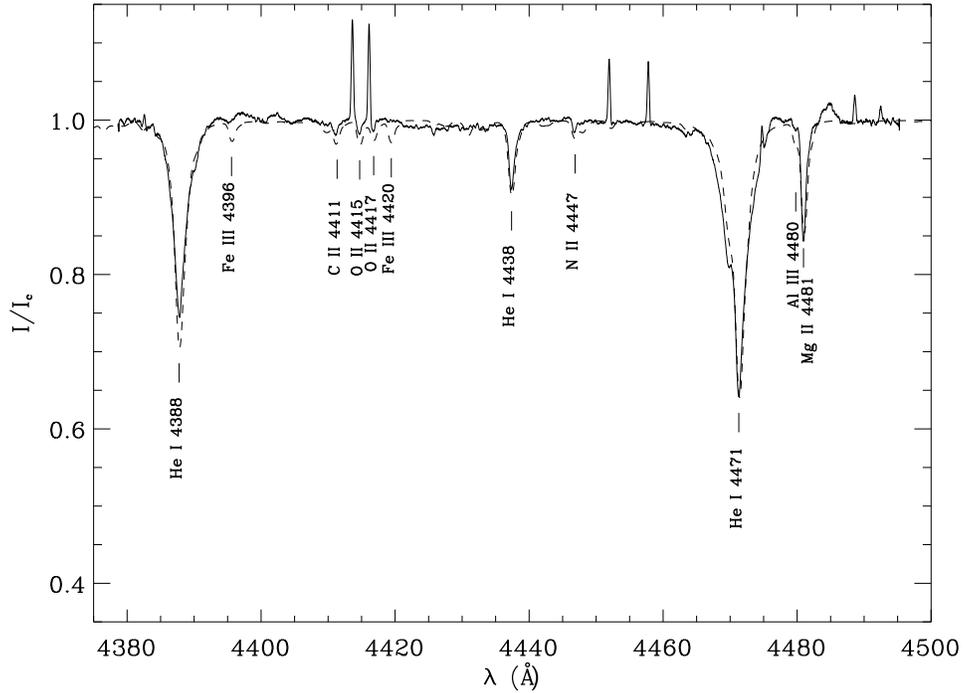}
\caption{ESPaDOnS spectrum (full line) of HD 200775 between 438 and 450 nm superimposed with the synthetic spectrum of a double star (dashed-line) calculated using TLUSTY non-LTE atmosphere models and the SYNSPEC code \citep{hubeny88,hubeny92,hubeny95}.}
\label{obssyn}
\end{figure*}

\begin{table*}
\caption{Fundamental parameters of HD 200775A and HD 200775B. Column 7 gives the age of the system.}
\label{fp}
\centering
\begin{tabular}{llllllll}
\hline\hline
Star   & $T_{\rm eff}$ & $\log(L/L_{\odot})$ & $M/M_{\odot}$ & $R/R_{\odot}$ & $\log g$          & age   & ${\rm v}\sin i$         \\
       & (K)           &                    &               &                            & (cm.s$^{\rm -2}$) &  (Myr) & (km.s$^{\rm -1}$) \\
\hline
HD 200775A & $18600 \pm 2000$ & $3.95 \pm 0.30$ & $10.7 \pm 2.5$ & $10.4 \pm 4.9$ & $3.4 \pm 1.2$ & \multirow{2}{1.5cm}{$0.1 \pm 0.05$} & $26 \pm 2$ \\
HD 200775B & $18600 \pm 2000$ & $3.77 \pm 0.30$ & $9.3 \pm 2.1$ & $  8.3 \pm 3.9$ & $3.6 \pm 1.2$& & $59 \pm 5$ \\
\hline
\end{tabular}
\end{table*}

We used the fundamental parameters of HD 200775 of \citet{hernandez04}. These authors studied the spectra of 75 stars, most of which are classified as Herbig Ae/Be stars. They determined the spectral type of each star as well as the luminosity with particular attention to the total-to-selective extinction ($R_V$) used. In the case of classical main sequence stars the interstellar extinction law gives $R_V=3.1$. However, the circumstellar matter of Herbig stars may be dominated by dust and can lead to a higher reddening of the star. Using $UBVR$ photometric data of \citet{herbst99}, \citet{hernandez04} found that $R_V=5$ fits better most stars of their sample than the standard value $R_V=3.1$. The knowledge of $R_V$ is fundamental to determine the luminosity: using $R_V=3.1$, they found $\log(L/L_{\odot})=3.73 \pm 0.27$, while with $R_V=5$, they found $\log(L/L_{\odot})=4.17 \pm 0.27$. For the purposes of this analysis, we adopt the most likely value of $R_V=5$, as HD 200775, illuminating the reflection nebulae NGC 7023, is still largely surrounded by dust and gas \citep{fuente98}, which certainly have an impact on the extinction law for this star.

The observed luminosity, derived by \citet{hernandez04}, corresponds the total of the luminosities of both stars. A priori, we do not know the luminosity ratio of the components. However, we can see in the spectrum that the lines of the secondary component are less deep than those of the primary. We suppose that this is mainly due to its higher rotational velocity (${\rm v}\sin i=59$ km.s$^{\rm -1}$, Section 3) and slightly due to a fainter luminosity. Furthermore, all lines observed in the spectrum of  the secondary are also observed in the spectrum of the primary star. For all these reasons, we suspect that the temperature of the secondary star is similar to that of the primary component. 

\citet{hernandez04} determined the spectral type of HD~200775 mainly by comparing the strength of atomic absorption lines to those of standard stars and found $\log(T_{\rm eff})=4.27$. We compared our spectra with a synthetic spectrum of a double star whose components have the same effective temperature and surface gravity :  $T_{\rm eff}=19000$ K and $\log g=3.5$. First, we calculated separate spectra of both components using TLUSTY non-LTE atmosphere models and the SYNSPEC code \citep{hubeny88,hubeny92,hubeny95}. Then we computed the spectrum of the double star, with the BINMAG1 program (O. Kochukhov, private communication), using a  macroturbulent velocity of 15 km/s, our measured ${\rm v}\sin i$, and radial velocities, and our estimation of the luminosity ratio $L_{\rm S}/L_{\rm P}=0.67$ (see below). We conclude that the temperatures of the synthetic spectrum are in good agreement with the observed spectrum of HD 200775. Figure \ref{obssyn} shows a portion of the observed spectrum superimposed on the synthetic one.

In addition, we also compared our spectra with a synthetic spectrum of a single star of $T_{\rm eff}=19000$ K and $\log g=3.5$. We found that the synthetic spectrum of a double star fits better the depth and the width of the lines than the spectrum of a single star. That adds additional support to our hypothesis that both components of the system have similar temperature. In order to quantify the temperature range of the stars, we have calculated synthetic spectra by varying the effective temperature of both stars around 19000K. We found that these spectra are able to fit our observations within $\pm2000$ K around 19000K.

The luminosity of \citet{hernandez04} is the sum of the luminosity of the primary ($L_{\rm P}$) and the secondary ($L_{\rm S}$) components of the system. We can estimate the luminosity ratio $r_{\rm L}=L_{\rm S}/L_{\rm P}$ from the Stokes $I$ profiles (normalised to the continuum of the binary). Section 3 describes the fitting procedure of these profiles. The result of the fitting procedure is the function:
\begin{equation}
I=1-f'_{\rm P}-f'_{\rm S}=1-\frac{1}{1+r_{\rm L}}f_{\rm P}-\frac{r_{\rm L}}{1+r_{\rm L}}f_{\rm S},
\end{equation}
where $f_{\rm P}$ and $f_{\rm S}$ are the separate Stokes $I$ profile shapes of the primary and the secondary stars. The ratio $\int f'_{\rm S}dv/\int f'_{\rm P}dv$ is therefore equal to $r_{\rm L} *W_{\rm S}/W_{\rm P}$, where $W_{\rm P}$ and $W_{\rm S}$ are the equivalent widths of the separate Stokes $I$ profiles of both stars. These profiles have been computed using the same mask, and therefore the same lines. Assuming a common origin for both stars of the system and that they are not peculiar, they should have approximately the same chemical composition. The intrinsic equivalent widths should therefore be identical and the ratio $\int f'_{\rm S}dv/\int f'_{\rm P}dv$ be equal to $r_{\rm L}$. We computed this ratio for all our observations. The mean and the standard deviation of these 30 values gives a luminosity ratio $L_{\rm S}/L_{\rm P}$ equal to $0.67 \pm 0.05$. We therefore derive the luminosity of the stars using : 
\begin{eqnarray}
	\log\frac{L_{\rm P}}{L_{\odot}} &=& \log\frac{L}{L_{\odot}}-\log(1+r_{\rm L}) = 3.95 \pm 0.30 \\
	\log\frac{L_{\rm S}}{L_{\odot}} &=& \log\frac{L}{L_{\odot}}-\log(1+\frac{1}{r_{\rm L}}) = 3.77 \pm 0.30
\end{eqnarray}
where, $L=L_{\rm P}+L_{\rm S}$ is the observed luminosity of HD 200775 \citep{hernandez04}.

\begin{figure}
\centering
\includegraphics[height=5cm,width=8cm]{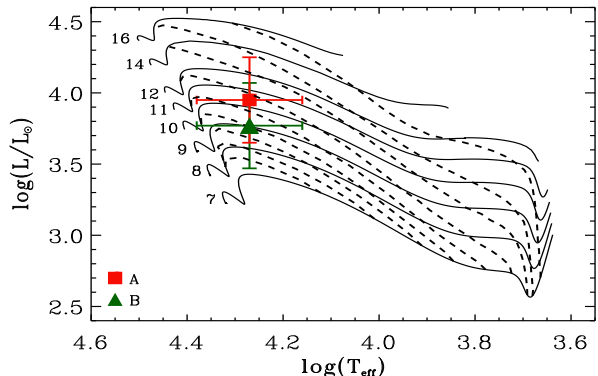}
\caption{Evolutionnary tracks (full lines) of different mass : from 7 M$_{\odot}$ to 16 M$_{\odot}$, plotted in a HR diagram. The dashed lines are the 0.01, 0.025, 0.04 Myr, 0.065 Myr, 0.09 Myr, 0.115 Myr, 0.14 Myr, 0.17 and 0.2 Myr isochrones. Crosses represent the error bars in temperature and luminosity of HD 200775A (red square) and HD 200775B (green triangle).}
\label{hr}
\end{figure}

We used the derived luminosities and the temperature of HD 200775 to place both stars in the HR diagram and to compare their position with evolutionary tracks of different masses calculated with the CESAM stellar evolutionary code \citep{morel97}. In this way we obtained the mass and the radius of both stars, as well as the age of the system, assuming a common origin for both stars (the age of the system is the intersection of both age ranges determined for each star separatly) (Fig. \ref{hr}). The value of the fundamental parameters are summarized in Table \ref{fp}. The mass ratio is therefore $q=1.1 \pm 0.5$, which is consistent with the value determined from the orbit analysis ($q=0.81\pm 0.22$, Sec. 3). Note that in both cases the error bars on the mass ratio are very large. This is partly due to our poor coverage of the orbital phase : our data cover only the half of the orbital period. In particular we do not have data where the maximum radial velocities are predicted. Therefore we need to get more data during the next years to improve it. The large error bars of the orbital parameters are also due to the non-unique solution of the fit of the Stokes $I$ profiles, leading to very large error bars on the radial velocities and the luminosity ratio. To better constrain the mass of the system, photometric observations of both separate components would be necessary. It would provide us the luminosity of each star and therefore we would no longer be dependent on our estimation of the luminosity ratio.

Furthermore, we note that using the orbit analysis we find that the secondary is more massive than the primary, whereas in this analysis we find that it is less massive than the primary. This inconsistency could result from an inaccurate estimation of the luminosity of both components, determined by assuming a similar reddening for both stars. This uncertainty, which we are unable to take into account in our error bars, could lead to an underestimation of the luminosity of the secondary relative to the primary and therefore to an overestimation of the mass of the secondary relative to the primary. This point is discussed in detail in Sec. 7.


\section{Properties of the intensity spectrum}

\subsection{Abundances anomalies}

Magnetic intermediate mass stars on the main sequence show strong photospheric abundance peculiarities. As some Herbig Ae/Be stars must be the evolutionary progenitors of these chemically peculiar Ap/Bp stars, we searched for abundances anomalies in our spectra, as well as time profile variations due to abundance spots. We chose in the same spectral region lines of different chemical species whose depths predicted by the synthetic spectra are identical. Then in the observed spectra, we compared the depths of these lines. 
We used this method, insensitive to the veiling phenomenon, to try to find over- or under-abundant species. We found no systematic differences between observed and calculated equivalent widths of numerous chemical species. We conclude that there are no strong abundance anomalies, in the limit of $\pm 0.40$ dex in abundance, in the spectrum of HD 200775. In particular, we observe no peculiarity in the helium lines nor any variability of the equivalent width of various species from one night to another and over many nights from Sept 2004 to April 2007.

These results are different from the behavior of most of the hot magnetic B stars on the main sequence, which are frequently He-rich or He-weak, and which show abundances starspots varying with the rotational phase of the star. {This suggests that at the young age of HD 200775, abundance anomalies have not had sufficient time to develop, or that they are limited by ongoing mixing due to accretion and mass loss.}


\subsection{The emission lines}

We observe two kinds of emission lines in the spectrum of HD~200775 : broad lines and narrow lines. Amongst the broad emission lines we find mainly the Balmer lines and the OI 7772~\AA $\;$ lines, and some Fe II and Si II with FWHM around 130 km.s$^{-1}$. The quality of our data as well as the complex structure of the emission lines do not allow us to fit them with a gaussian function, and therefore to measure the radial velocities of these lines. However the bisector velocities of these lines follow the radial velocity of the secondary (Sec. 3). We note a decrease of the equivalent widths of these emission lines from Sept 2004 to April 2007 (see Fig. \ref{fitvrad} upper panel, for H$\alpha$). We observe in Fig. \ref{fitvrad} that the maximum equivalent width of H$\alpha$ occurs close to the periastron passage of the orbit (indicated by the arrows) that we derive in Section 3. This is consistent with the hypothesis of \citet{pogodin04} who suggest that binarity could be the origin of these variations.

In addition to these broad lines, we note a large number of narrow emission lines, about a hundred over the whole spectrum, with a FWHM around 19 km.s$^{-1}$. Some of these are clearly visible in Fig. \ref{obssyn}. {We do not observe any shift in radial velocity of these lines during our 2-year observations.} Using the numerous observations of $\eta$ Carinae \citep{thackeray53,thackeray62,thackeray67,hamann94} we were able to identify all of them and to conclude that they are nebular lines (lines coming from the nebulosity illuminated by HD 200775). Appendix A gives the identification of all lines.

%

\section{Magnetic field analysis}

\begin{figure}
\centering
\includegraphics[height=5.5cm]{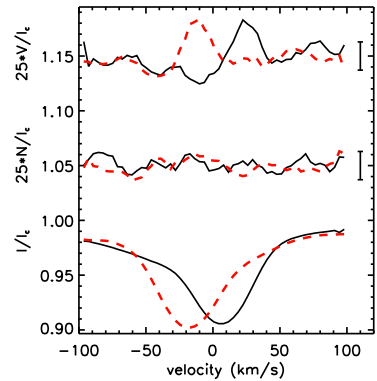}
\caption{Stokes $I$ (bottom) and $V$ (top) mean profiles of HD 200775 observed on Sept. 23rd 2004 (full line) and May 23rd 2005 (dashed line). The null profile $N$ is plotted in the middle, and the 3$\sigma$ error bars of $V$ and $N$ are plotted next to the respective profiles.}
\label{maisept}
\end{figure}

Figure \ref{maisept} shows the Stokes $I$ and $V$ LSD profiles obtained on Sept. 2004 and May 23rd 2005. First we note that the null profile $N$ (Sec. 2) is totally flat indicating that the signature in the $V$ profile is real. Secondly the $V$ profile has shifted from the blue to the red, as the primary component of the $I$ profile, between Sept. 2004 and May 2005. This behaviour is observed throughout the entire data set. For this reason we attribute the observed magnetic field solely to the primary component.

To investigate the possible presence of a magnetic field in the secondary, we calculated synthetic Stokes $V$ profiles for a star similar to the secondary one, with a ${\rm v}\sin i$ of 59~km.s$^{-1}$, and considering the same magnetic proprieties of the primary, described in Section 6.2. The maximum signal in circular polarisation predicted is around 10$^{-3}$, which is the same order of magnitude as the error bars of our observed $V$ profiles. Therefore, if the secondary component of the system hosts a magnetic field of similar topology and similar intensity to that of the primary, it would be difficult to detect it with our data. On the other hand, if our polarisation spectra contain a signal of the secondary star corresponding to a similar field intensity,  it is negligible with respect to the signal of the primary star.

\subsection{Phased longitudinal field variation}

Figure \ref{fitallv} shows the variations of the 30 Stokes $V$ profiles, normalised to the continuum intensity of the binary, observed from 2004 to 2007. The longitudinal magnetic field $B_{\ell}$, which is the projected magnetic field onto the line of sight, can be obtained from both LSD Stokes $I$ and $V$ profiles, as explained by \citet{donati97} and \citet{wade00}. With this aim we compute the Stokes $I$ and $V$ profiles of the primary ($I_{\rm P}$ and $V_{\rm P}$) as follows. Using Eq. (1) and the fit of the Stokes $I$ profiles of the binary (Sec. 3), we subtract from the observed Stokes $I$ profile the synthetic profile of the secondary component. Then we renormalised to the continuum of the primary:
\begin{equation}
I_{\rm P} = 1-(1+r_{\rm L})(1-I-f'_{\rm S}).
\end{equation}
As we assume that the observed magnetic field is only from the primary, we only need to renormalise the observed Stokes $V$ profile by the continuum of the primary, in order to get $V_{\rm P}$:
\begin{equation}
V_{\rm P} = (1+r_{\rm L})V
\end{equation}

\begin{figure}
\centering
\includegraphics[height=8cm, angle=90]{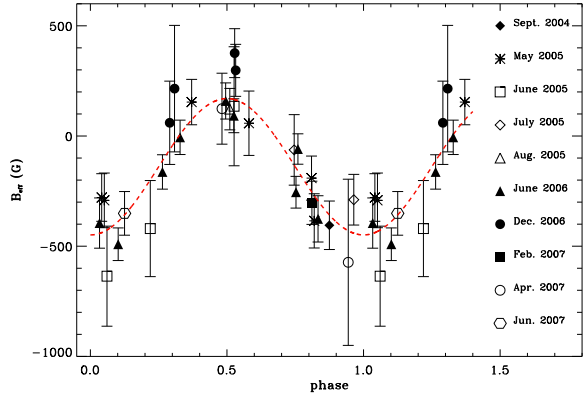}
\caption{Longitudinal magnetic field plotted in function of the rotation phase. The dashed line is the best sinusoidal fit.}
\label{bphaseall}
\end{figure}

Using these new profiles and Eq. (1) of \citet{wade00}, we measured the longitudinal magnetic field $B_{\ell}$ of our data. We fit the variations of $B_{\ell}$ with a sinusoidal function of 4 parameters : 
\begin{itemize}
	\item $P$, the rotation period of the star, 
	\item $t_0$, the reference Julian Date of the maximum of $B_{\ell}$, 
	\item $B$, the semi-amplitude of the curve, 
	\item $B_0$, the shift of the sinusoidal curve with respect to $B_{\ell}=0$,
\end{itemize}
consistent with a dipole centered inside the star. The best fit superimposed on the data in Fig. \ref{bphaseall}, gives the following parameters : $P=4.328 \pm 0.003$ d, $t_0=2453515.8 \pm 0.3$ d,  $B=-309 \pm 115$ G and $B_0=-139 \pm 95$ G, with a reduced $\chi^2=1.1$. We used that value as a first estimation of the rotation period in the fitting procedure of the Stokes $V$ profiles.

This analysis reveals the basic properties of the magnetic field of HD~200775A: that it is organised on large scales; that it has an important global dipole component; and that the dipole geometry is stable on timescales of several years.

\subsection{Fitting of Stokes V profiles}

\begin{figure*}
\centering
\includegraphics[angle=90,width=17cm]{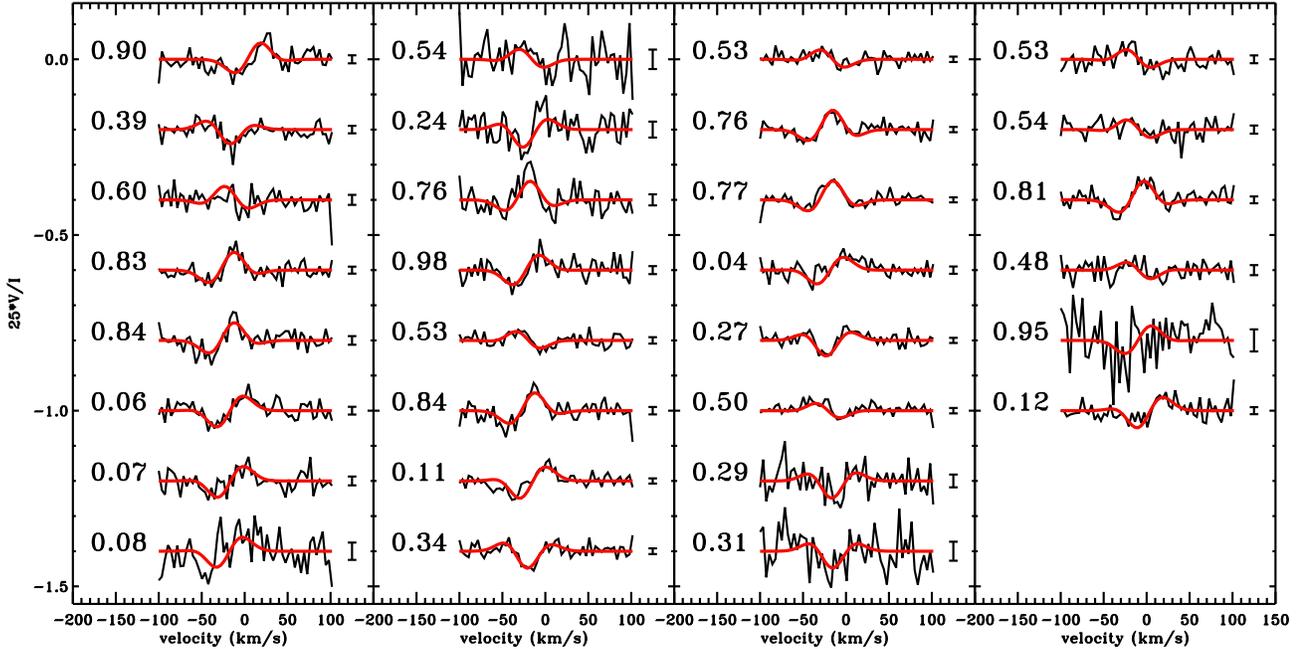}
\caption{LSD $V$ profiles (noisy black line) of the 30 spectra superimposed to the best oblique rotator model (smooth red line). The numbers close to the profiles are the rotation phase, and the little bars on the right of the profiles are the mean error bars in $V$. The profiles are sorted by increasing date, as in Fig. 1.}
\label{fitallv}
\end{figure*}

To fit the Stokes $V$ profiles directly, we use the decentred dipole oblique rotator model described by \citet{landstreet70}. Using the relations of \citet{landstreet70}, giving the intensity of the magnetic field inside a star, we calculate the longitudinal magnetic field $b_{\ell}(\theta,\varphi)$, at each point ($\theta$,$\varphi$) of the surface of the star of velocity $\rm v$, in classical spherical coordinates in the observer's frame.

We assume a gaussian local intensity profile of width $\sigma$ and depth $d$. The width is calculated using the resolving power of the instrument and the macroturbulent velocity determined in Section 3. The depth is determined by fitting the Stokes $I$ profiles of the primary component, determined in Section 6.1. We then calculated the local Stokes $V$ profile at each point on the surface of the star using the weak magnetic field relation of \citet{landi73} : 
\begin{equation}
	V({\rm v},\theta,\varphi)=-C\overline{g}\lambda_0cb_{\ell}(\theta,\varphi)\frac{{\rm d}I}{{\rm dv}},
\end{equation}
where $C=4.67\times10^{-13}$\AA$^{-1}$G$^{-1}$, $\overline{g}$ and $\lambda_0$ (\AA) are the mean Land{\'e} factor and wavelength of the lines used in the mask (Section 2), $c$ is the speed of light, $I$ is the local intensity profile, and $\rm v$ is the velocity. Then we integrated over the visible stellar surface using the limb darkening law with a parameter equal to 0.4 \citep{claret00}. We obtain the synthetic Stokes $V({\rm v})$ profile, that we normalised to the intensity continuum, to compare to the observed $V$ profiles. This model depends on five parameters : 
\begin{itemize}
	\item $P$ the rotation period
	\item $t_0$, the reference Julian Date of the maximum of the surface magnetic intensity, used with $P$ to compute the rotation phase
	\item $i$, the inclination of the stellar rotation axis to the observers line-of-sight,
	\item $\beta$ the magnetic obliquity angle, 
	\item $B_{\rm d}$ the dipole magnetic intensity,
	\item $d_{\rm dip}$ the displacement of the dipole from the center of the star, along the magnetic axis, in stellar radii ($R_*$).
\end{itemize}

\begin{table}
\caption{Ranges and minimum bins of parameters explored in the fit of the Stokes $V$ profiles}
\label{tab:grid}
\centering
\begin{minipage}[t]{\linewidth}
\begin{tabular}{llll}
\hline\hline
Parameters & Min & Max & bin \\
\hline
P (days) & 4.0 & 5.0 & 0.0001\\
$T_0$ (HJD) & 2453514 &  2453519 & 0.02\\
$i$ ($^{\circ}$) & 0 & 90 & 1\\
$\beta$ ($^{\circ}$) & 0 & 180 & 1\\
$B_{\rm d}$ (G) & 0 & 2000 & 10\\
$d_{\rm dip}$ ($R_*$) & -0.4 & 0.4 & 0.01\\
\hline
\end{tabular}
\end{minipage}
\end{table}

We calculated a grid of $V$ profiles for each date of observation (see Table \ref{tab:grid} for details on the grid), varying the six parameters (and assuming for the initial value of $P$ the solution obtained from modelling the longitudinal field variation). Then we applied a $\chi^2$ minimisation to find the best model which matches simultaneously our 30 observed profiles. The best model that we found, with $\chi^2=1.0$, corresponds to $P=4.3281 \pm 0.0010$ d, $i=60\pm11^{\circ}$, $\beta=125 \pm 8^{\circ}$, $B_{\rm d}=1000 \pm 150$ G, and $d_{dip}=0.05 \pm 0.04$~$R_*$, where the error bars correspond to 3$\sigma$ confidence. Figure \ref{fitallv} shows the synthetic Stokes $V$ profiles superimposed on the observed ones. We see that this model acceptably reproduces most of the observed $V$ profiles.
The small value of the dipole decentring parameter is nearly consistent with zero. The available data are therefore consistent with a dipolar magnetic field at the centre of the star

In order to compare this result to the fit of the longitudinal field variations, we estimate the intensity $B_{\rm d}$ and the obliquity angle $\beta$ of the magnetic dipole component using the fit of the longitudinal field variations. Using the well-known equations relating $i$, $\beta$ and $B_{\rm d}$ to the longitudinal field extrema \citep{borra80}, and adopting $i=60\pm11^{\circ}$, we find $\beta=128 \pm 13^{\circ}$ and $B_{\rm d}=1500^{+1100}_{-700}$ G. These values agree within the error bars with those determined with the fit of the Stokes $V$ parameters, which confirms the topology of the magnetic field found using this fit. Table \ref{tab:mg} summarises the parameters of the adopted dipole model.

Because the Stokes $I$ and $V$ profiles are influenced in similar ways by uncertainties in the procedure that we have used to recover the individual LSD profiles of the primary, the magnetic modeling is only weakly sensitive to the details of the procedures described in Sect. 4 and Sect. 6. From a variety of experiments we find that neither the assumed luminosity ratio, nor the details of the fitting procedure yielding the ${\rm v}\sin i$s, profile depths and radial velocities, have any significant direct influence on the derived magnetic models.

Finally, we point out that the magnetic modelling provides a strong constraint on the inclination of the rotation axis $i$. In particular, for smaller values of $i$, we are unable to reproduce the Stokes $V$ profile shapes.

\begin{table}
\caption{Magnetic dipole model of HD~200775A}
\label{tab:mg}
\centering
\begin{minipage}[t]{\linewidth}
\begin{tabular}{ll}
\hline\hline
P (days) & $4.3281 \pm 0.0010$ \\
$T_0$ (HJD) & $2453515.8 \pm 0.8$ \\
$i$ ($^{\circ}$) & $60 \pm 11$ \\
$\beta$ ($^{\circ}$) & $ 125 \pm 8$ \\
$B_{\rm d}$ (G) & $1000 \pm 150$ \\
$d_{\rm dip}$ ($R_*$) & $0.05 \pm 0.04$ \\
\hline
\end{tabular}
\end{minipage}
\end{table}


\section{Discussion}

\subsection{Consequences of the magnetic model on the fundamental parameters of the star}

Using our determinations of the ${\rm v}\sin i$ (from the LSD $I$ profiles) and the period and inclination (from the magnetic model), we can estimate the radius of HD~200775A. We find $R=2.6\pm0.5 R_{\odot}$, which is significantly smaller than the radius found from stellar evolutionary models (Table \ref{fp}). In particular, this value suggest a star close to the main sequence, contrary to the stellar models. The model value is determined from the stellar effective temperature (which we consider to be well-determined) and the inferred luminosity of the primary star. The primary's luminosity is itself determined based on the ratio of equivalent widths of the LSD profiles, and from the dereddened photometry of the combined system supported by \cite{hernandez04}, which assumed equal reddening for each of the two components.

The inferred luminosities could be affected by two possibly incorrect assumptions inherent in this analysis. First, based on our derivation of nearly identical effective temperatures for the two components, we have assumed that the intrinsic equivalent widths of their LSD profiles are equal. We then used the observed equivalent widths of the two components, determined from the LSD profiles, to infer their relative luminosities. However, if the temperatures of the primary and secondary differ by the maximum allowed by the error bars (about 4000 K), the combined effects of differences in excitation and ioniosation, and the $T_{\rm eff}^4$ dependence of the flux contribution, could combine to generate important relative differences (either larger or smaller depending on the effective temperatures) in the contributions of the spectra of the two components to the observed LSD profiles.

The second assumption that could influence the inferred luminosities is that the two components are affected by similar reddening. In their interferometric study of the HD 200775 system, \citet{monnier06}\ resolved the circumstellar disc of the secondary and found that the secondary was brighter than the primary in H band by a factor of 6.5. If this IR flux difference results from the presence of CS material around the secondary that is not present around the primary (which would be consistent with our spectroscopic observations, indicating that the H$\alpha$ emission is associated with the secondary), it would imply that the secondary is affected by additional reddening that has not been taken into account in our analysis. The increased extinction of the secondary would result in an underestimated luminosity of this star relative to the primary. Consequently, the primary's luminosity would be overestimated, leading to a systematic overestimation of its inferred radius. It would also lead to an overestimation of the mass of the primary relative to the secondary and therefore to a mass ratio $q=M_{\rm P}/M_{\rm S}$ greater than one (as observed), which is inconsistent with the orbit determination of $q=0.81\pm 0.22$.

Neither of these potentially important sources of systematic error can be investigated in detail with the available data, and ultimately we are led to conclude that the primary's radius could be anywhere from the ZAMS radius to the radius reported in Table~3. Likewise the masses and ages of the stars derived in Sec. 4 from the luminosity and temperature must be considered with caution. Individual photometric observations of each of the stars is required to get independent measurements of the luminosities of both stars, and therefore accurate values of their masses, radii and ages.


\subsection{Consequences of the magnetic model on the origin of the system}

Assuming $i=60^{\circ}$  implies that the primary would be close to the main sequence, and therefore would have a lower luminosity than the secondary. According to the luminosity ratio in H band of Monnier et al. (2006), the secondary is redder that the primary and therefore seems to be surrounded by denser circumstellar matter than the secondary, which is consistent with our spectroscopic data. This would indicate that the secondary is in a younger evolutionary state than the primary. Furthermore the eccentricity of the orbit found in Sec. 3 is very large (Table 2). These properties may well suggest that the two components are at two very different evolutionary stages, which could imply that the system formed by capture instead of forming initially as a simple close binary.


\section{Conclusions}

In the framework of the understanding of the origin of the magnetic field of the intermediate and high mass stars, we have begun to acquire spectropolarimetric observations of their potential progenitors, the Herbig Ae/Be stars, including HD 200775. A magnetic field has been discovered in the latter and in order to determine its topology, we monitored it during many successive nights and over more than 2 years, using the high resolution spectropolarimeters ESPaDOnS and Narval.

First we inspected the intensity spectrum and found that this star is a double-lined spectroscopic binary, whose secondary seems similar, in temperature and luminosity, to the primary. We measured accurate values of the projected rotational velocity of both stars: $26\pm2$ km.s$^{-1}$ and $59\pm5$ km.s$^{-1}$. We fitted the radial velocity curves of both components and found an ephemeris similar to that of Pogodin et al. (2004). The amplitude of the radial velocity curves leads to a poorly constrained mass ratio. Observations of the separate components, in order to better determine the luminosity ratio of the system, are required to determine accurately the luminosities and hence the masses of the components. According to our determination of the orbital period and of the masses of both stars, we estimate the separation between both components of the system around 16 mas (the distance of the system is estimated around 430 pc by \citet{ancker97}). Only interferometric observations of the system would be able to separate the two components and to estimate the luminosity ratio. 

We have also shown that the broad emission observed in the intensity spectrum is linked to the secondary component of the system. We found no evidence of abundance anomalies in the spectrum. Finally, many narrow emission lines have been observed over the whole spectrum, which we identify as nebular lines.

The position and velocity variations of the Stokes $V$ profile shows that the detected Zeeman signature corresponds solely to the primary component. However, if the secondary hosts a magnetic field of similar topology and intensity to the primary, the signal in polarisation would be negligible with respect to the signal of the primary. On the other hand, this implies that our observations are relatively insensitive to the presence of a field in the secondary.

We modeled the temporal variations of the Stokes $V$ profiles in two different ways. First we measured the longitudinal magnetic field of the star and we fitted a sinusoidal curve, as predicted by the dipole oblique rotator model. Then we used the period found as a first estimation of the rotation period of the star in the Stokes $V$ profiles fitting procedure. We considered an oblique rotator model and we applied $\chi^2$ minimisation to match simultaneously the 30 Stokes $V$ profiles that we observed over more than one year. We find that this simple decentered dipole model is sufficient to fit most of the profiles. In this way we provide the magnetic field topology: a dipole of intensity $\sim 1000$ G, displaced by $0.05 \pm 0.04$~$R_*$ from the center of the star towards the positive magnetic pole, whose the rotation axis is inclined by $60 \pm 11^{\circ}$ with respect to the observer line-of-sight, and magnetic axis is inclined by $125 \pm 8 ^{\circ}$ with respect to the rotation axis. HD 200775A rotates with a period of $4.3281 \pm 0.0010$ d. 

We therefore show that the magnetic field of this star is approximately dipolar, strongly inclined with respect to the rotation axis, with a polar intensity of 1000 G, and stable over more than two years. These characteristics are similar to the topology of the magnetic fields of the Ap/Bp stars \citep{borra80,bohlender87}.

We have concluded that significant uncertainties exist related to the luminosity, and therefore the mass, radius and age of the magnetic primary star. To clarify these uncertainties and to move forward, additional photometric observations {\em of the individual components} are required. In addition, a more complete spectroscopic coverage of the orbital cycle is needed, with the hope of ultimately performing spectral disentangling to yield the individual spectra of the primary and secondary.

%

\section*{Acknowledgments}

We warmly thank Georges Alecian for fruitful discussions. We are very grateful to O. Kochukhov, who provided his BINMAG1 code, and who tried to apply his spectral disentangling method to our data. We thank G. Mathys, the referee, for his judicious comments, which led to major improvements in the paper. EA is supported by the Marie Curie FP6 program. GAW acknowledges support from the Natural Science and Engineering Research Council of Canada (NSERC) and the DND Academic Research Programme (ARP).

%

\nocite{*}
\bibliographystyle{mn2e}
\bibliography{hd200775}

\begin{thebibliography}{}

\bibitem[\protect\citeauthoryear{{Bohlender}, {Landstreet}, {Brown} \&
  {Thompson}}{{Bohlender} et~al.}{1987}]{bohlender87}
{Bohlender} D.~A.,  {Landstreet} J.~D.,  {Brown} D.~N.,    {Thompson} I.~B.,
  1987, ApJ, 323, 325

\bibitem[\protect\citeauthoryear{{Borra} \& {Landstreet}}{{Borra} \&
  {Landstreet}}{1980}]{borra80}
{Borra} E.~F.,  {Landstreet} J.~D.,  1980, ApJS, 42, 421

\bibitem[\protect\citeauthoryear{{Catala}, {Alecian}, {Donati}, {Wade},
  {Landstreet}, {B{\"o}hm}, {Bouret}, {Bagnulo}, {Folsom} \&
  {Silvester}}{{Catala} et~al.}{2007}]{catala07}
{Catala} C.,  {Alecian} E.,  {Donati} J.-F.,  {Wade} G.~A.,  {Landstreet}
  J.~D.,  {B{\"o}hm} T.,  {Bouret} J.-C.,  {Bagnulo} S.,  {Folsom} C.,
  {Silvester} J.,  2007, A\&A, 462, 293

\bibitem[\protect\citeauthoryear{{Catala}, {Bohm}, {Donati} \&
  {Semel}}{{Catala} et~al.}{1993}]{catala93}
{Catala} C.,  {Bohm} T.,  {Donati} J.-F.,    {Semel} M.,  1993, A\&A, 278, 187

\bibitem[\protect\citeauthoryear{{Catala}, {Donati}, {B{\"o}hm}, {Landstreet},
  {Henrichs}, {Unruh}, {Hao}, {Collier Cameron} \& et al.}{{Catala}
  et~al.}{1999}]{catala99}
{Catala} C.,  {Donati} J.~F.,  {B{\"o}hm} T.,  {Landstreet} J.,  {Henrichs}
  H.~F.,  {Unruh} Y.,  {Hao} J.,  {Collier Cameron} A.,    et al. 1999, A\&A,
  345, 884

\bibitem[\protect\citeauthoryear{{Claret}}{{Claret}}{2000}]{claret00}
{Claret} A.,  2000, A\&A, 363, 1081

\bibitem[\protect\citeauthoryear{{Corporon} \& {Lagrange}}{{Corporon} \&
  {Lagrange}}{1999}]{corporon99}
{Corporon} P.,  {Lagrange} A.-M.,  1999, A\&AS, 136, 429

\bibitem[\protect\citeauthoryear{{Donati}, {Semel}, {Carter}, {Rees} \&
  {Collier Cameron}}{{Donati} et~al.}{1997}]{donati97}
{Donati} J.-F.,  {Semel} M.,  {Carter} B.~D.,  {Rees} D.~E.,    {Collier
  Cameron} A.,  1997, MNRAS, 291, 658

\bibitem[\protect\citeauthoryear{{Fuente}, {Martin-Pintado}, {Rodriguez-Franco}
  \& {Moriarty-Schieven}}{{Fuente} et~al.}{1998}]{fuente98}
{Fuente} A.,  {Martin-Pintado} J.,  {Rodriguez-Franco} A.,
  {Moriarty-Schieven} G.~D.,  1998, A\&A, 339, 575

\bibitem[\protect\citeauthoryear{{Gray}}{{Gray}}{1992}]{gray92}
{Gray} D.~F.,  1992, {The observation and analysis of stellar photospheres}.
Cambridge Astrophysics Series, Cambridge: Cambridge University Press, 1992, 2nd
  ed., ISBN 0521403200.

\bibitem[\protect\citeauthoryear{{Hamann}, {Depoy}, {Johansson} \&
  {Elias}}{{Hamann} et~al.}{1994}]{hamann94}
{Hamann} F.,  {Depoy} D.~L.,  {Johansson} S.,    {Elias} J.,  1994, ApJ, 422,
  626

\bibitem[\protect\citeauthoryear{{Herbst} \& {Shevchenko}}{{Herbst} \&
  {Shevchenko}}{1999}]{herbst99}
{Herbst} W.,  {Shevchenko} V.~S.,  1999, AJ, 118, 1043

\bibitem[\protect\citeauthoryear{{Hern{\'a}ndez}, {Calvet}, {Brice{\~n}o},
  {Hartmann} \& {Berlind}}{{Hern{\'a}ndez} et~al.}{2004}]{hernandez04}
{Hern{\'a}ndez} J.,  {Calvet} N.,  {Brice{\~n}o} C.,  {Hartmann} L.,
  {Berlind} P.,  2004, AJ, 127, 1682

\bibitem[\protect\citeauthoryear{{Hubeny}}{{Hubeny}}{1988}]{hubeny88}
{Hubeny} I.,  1988, Comput. Phys. Comm., 52, 103

\bibitem[\protect\citeauthoryear{{Hubeny} \& {Lanz}}{{Hubeny} \&
  {Lanz}}{1992}]{hubeny92}
{Hubeny} I.,  {Lanz} T.,  1992, A\&A, 262, 501

\bibitem[\protect\citeauthoryear{{Hubeny} \& {Lanz}}{{Hubeny} \&
  {Lanz}}{1995}]{hubeny95}
{Hubeny} I.,  {Lanz} T.,  1995, ApJ, 439, 875

\bibitem[\protect\citeauthoryear{{Hubrig}, {Pogodin}, {Yudin}, {Sch{\"o}ller}
  \& {Schnerr}}{{Hubrig} et~al.}{2007}]{hubrig07}
{Hubrig} S.,  {Pogodin} M.~A.,  {Yudin} R.~V.,  {Sch{\"o}ller} M.,    {Schnerr}
  R.~S.,  2007, A\&A, 463, 1039

\bibitem[\protect\citeauthoryear{{Hubrig}, {Sch{\"o}ller} \& {Yudin}}{{Hubrig}
  et~al.}{2004}]{hubrig04}
{Hubrig} S.,  {Sch{\"o}ller} M.,    {Yudin} R.~V.,  2004, A\&A, 428, L1

\bibitem[\protect\citeauthoryear{{Kurucz}}{{Kurucz}}{1993}]{kurucz93}
{Kurucz} R.,  1993, Opacities for Stellar Atmospheres:
  [-3.5],[-4.0],[-4.5].~Kurucz CD-ROM No.~7.~Cambridge, Mass.: Smithsonian
  Astrophysical Observatory, 1993., 7

\bibitem[\protect\citeauthoryear{{Landi degl'Innocenti} \& {Landi
  degl'Innocenti}}{{Landi degl'Innocenti} \& {Landi
  degl'Innocenti}}{1973}]{landi73}
{Landi degl'Innocenti} E.,  {Landi degl'Innocenti} M.,  1973, Sol. Phys., 31,
  299

\bibitem[\protect\citeauthoryear{{Landstreet}}{{Landstreet}}{1970}]{landstreet%
70}
{Landstreet} J.~D.,  1970, ApJ, 159, 1001

\bibitem[\protect\citeauthoryear{{Miroshnichenko}, {Mulliss}, {Bjorkman},
  {Morrison}, {Glagolevskij} \& {Chountonov}}{{Miroshnichenko}
  et~al.}{1998}]{miroshnichenko98}
{Miroshnichenko} A.~S.,  {Mulliss} C.~L.,  {Bjorkman} K.~S.,  {Morrison} N.~D.,
   {Glagolevskij} Y.~V.,    {Chountonov} G.~A.,  1998, PASP, 110, 883

\bibitem[\protect\citeauthoryear{{Monnier}, {Berger}, {Millan-Gabet}, {Traub},
  {Schloerb}, {Pedretti}, {Benisty}, {Carleton}, {Haguenauer}, {Kern},
  {Labeye}, {Lacasse}, {Malbet}, {Perraut}, {Pearlman} \& {Zhao}}{{Monnier}
  et~al.}{2006}]{monnier06}
{Monnier} J.~D.,  {Berger} J.-P.,  {Millan-Gabet} R.,  {Traub} W.~A.,
  {Schloerb} F.~P.,  {Pedretti} E.,  {Benisty} M.,  {Carleton} N.~P.,
  {Haguenauer} P.,  {Kern} P.,  {Labeye} P.,  {Lacasse} M.~G.,  {Malbet} F.,
  {Perraut} K.,  {Pearlman} M.,    {Zhao} M.,  2006, ApJ, 647, 444

\bibitem[\protect\citeauthoryear{{Morel}}{{Morel}}{1997}]{morel97}
{Morel} P.,  1997, A\&AS, 124, 597

\bibitem[\protect\citeauthoryear{{Palla} \& {Stahler}}{{Palla} \&
  {Stahler}}{1993}]{palla93}
{Palla} F.,  {Stahler} S.~W.,  1993, ApJ, 418, 414

\bibitem[\protect\citeauthoryear{{Pogodin}, {Miroshnichenko}, {Tarasov},
  {Mitskevich}, {Chountonov}, {Klochkova}, {Yushkin}, {Manset}, {Bjorkman},
  {Morrison} \& {Wisniewski}}{{Pogodin} et~al.}{2004}]{pogodin04}
{Pogodin} M.~A.,  {Miroshnichenko} A.~S.,  {Tarasov} A.~E.,  {Mitskevich}
  M.~P.,  {Chountonov} G.~A.,  {Klochkova} V.~G.,  {Yushkin} M.~V.,  {Manset}
  N.,  {Bjorkman} K.~S.,  {Morrison} N.~D.,    {Wisniewski} J.~P.,  2004, A\&A,
  417, 715

\bibitem[\protect\citeauthoryear{{Stift}}{{Stift}}{1975}]{stift75}
{Stift} M.~J.,  1975, MNRAS, 172, 133

\bibitem[\protect\citeauthoryear{{Thackeray}}{{Thackeray}}{1953}]{thackeray53}
{Thackeray} A.~D.,  1953, MNRAS, 113, 211

\bibitem[\protect\citeauthoryear{{Thackeray}}{{Thackeray}}{1962}]{thackeray62}
{Thackeray} A.~D.,  1962, MNRAS, 124, 251

\bibitem[\protect\citeauthoryear{{Thackeray}}{{Thackeray}}{1967}]{thackeray67}
{Thackeray} A.~D.,  1967, MNRAS, 135, 51

\bibitem[\protect\citeauthoryear{{van den Ancker}, {The}, {Tjin A Djie},
  {Catala}, {de Winter}, {Blondel} \& {Waters}}{{van den Ancker}
  et~al.}{1997}]{ancker97}
{van den Ancker} M.~E.,  {The} P.~S.,  {Tjin A Djie} H.~R.~E.,  {Catala} C.,
  {de Winter} D.,  {Blondel} P.~F.~C.,    {Waters} L.~B.~F.~M.,  1997, A\&A,
  324, L33

\bibitem[\protect\citeauthoryear{{Wade}, {Bagnulo}, {Drouin}, {Landstreet} \&
  {Monin}}{{Wade} et~al.}{2007}]{wade07}
{Wade} G.~A.,  {Bagnulo} S.,  {Drouin} D.,  {Landstreet} J.~D.,    {Monin} D.,
  2007, MNRAS, 376, 1145

\bibitem[\protect\citeauthoryear{{Wade}, {Donati}, {Landstreet} \&
  {Shorlin}}{{Wade} et~al.}{2000}]{wade00}
{Wade} G.~A.,  {Donati} J.-F.,  {Landstreet} J.~D.,    {Shorlin} S.~L.~S.,
  2000, MNRAS, 313, 851

\bibitem[\protect\citeauthoryear{{Wade}, {Drouin}, {Bagnulo}, {Landstreet},
  {Mason}, {Silvester}, {Alecian}, {B{\"o}hm}, {Bouret}, {Catala} \&
  {Donati}}{{Wade} et~al.}{2005}]{wade05}
{Wade} G.~A.,  {Drouin} D.,  {Bagnulo} S.,  {Landstreet} J.~D.,  {Mason} E.,
  {Silvester} J.,  {Alecian} E.,  {B{\"o}hm} T.,  {Bouret} J.-C.,  {Catala} C.,
     {Donati} J.-F.,  2005, A\&A, 442, L31

\bibitem[\protect\citeauthoryear{{Whitcomb}, {Gatley}, {Hildebrand}, {Keene},
  {Sellgren} \& {Werner}}{{Whitcomb} et~al.}{1981}]{whitcomb81}
{Whitcomb} S.~E.,  {Gatley} I.,  {Hildebrand} R.~H.,  {Keene} J.,  {Sellgren}
  K.,    {Werner} M.~W.,  1981, ApJ, 246, 416

\end{thebibliography}

%

\appendix
\section{Nebular lines}
\onecolumn

\begin{longtable}{lrllrl}
\caption{\label{nebular} Identification of the nebular lines in the spectrum of HD 200775. Column 1 gives the wavelength observed in our spectra, column 2 gives the equivalent width of the line, column 3 gives the ion, column 4 gives the multiplet number, column 5 gives the rest wavelength of the line, and column 6 gives the reference. I: \citet{thackeray53}, II: \citet{thackeray62}, III: \citet{thackeray67}, IV: \citet{hamann94}.}\\
\hline
$\lambda$ (\AA) & $W_{\lambda}$ (m\AA) & Ion & Identification & $\lambda_0$ (\AA) & Réf. \\ \hline
\hline
\endfirsthead
\caption{continued.}\\
\hline\hline
$\lambda$ (\AA) & $W_{\lambda}$ (m\AA) & Ion & Identification & $\lambda_0$ (\AA) & Réf. \\ \hline
\hline
\endhead
\hline
\endfoot
4114.34 & 8.5     & Fe II & 23F & 4114.48 & III \\
4177.07 & 10.9   & Fe II & 21F & 4177.21 & III \\
4210.99 & PCI \footnote{Inverse Pcygni profile, the equivalent width have therefore not been determined}    & Fe II & 23F & 4211.10 & III \\
4243.83 & 42.6   & Fe II & 21F & 4243.98 & III \\
4244.68 & 11.7   & Fe II & 21F & 4244.81 & III \\
4276.69 & 40.7   & Fe II & 21F & 4276.83 & I   \\
4287.25 & 85.5   & Fe II & 7F   & 4287.40 & I   \\
4305.75 & 9.3     & Fe II & 21F & 4305.90 & I   \\
4319.47 & 19.7   & Fe II & 21F & 4319.62 & I   \\
4326.13 & 5.4     & Ni  II & 3F   & 4326.28 & I   \\
4346.72 & 10.0   & Fe II & 21F & 4246.85 & I   \\
4352.64 & 19.5   & Fe II & 21F & 4352.78 & I   \\
4358.23 & 22.9   & Fe II & 21F & 4358.37 & I   \\
4359.18 & 70.6   & Fe II & 7F   & 4359.34 & I   \\
4372.28 & 9.4     & Fe II & 21F & 4372.43 & I   \\
4382.57 & 5.2     & Fe II & 6F   & 4382.75 & I   \\
4413.62 & 56.4   & Fe II & 7F   & 4413.78 & I   \\
4416.11 & 51.1   & Fe II & 6F   & 4416.27 & I   \\
4451.95 & 30.2   & Fe II & 7F   & 4452.95 & I   \\
4457.80 & 29.7   & Fe II & 6F   & 4457.95 & I   \\
4474.75 & 20.0   & Fe II & 7F   & 4474.91 & I   \\
4488.60 & 12.1   & Fe II & 6F   & 4488.75 & I   \\
4492.49 & 7.6     & Fe II & 6F   & 4492.64 & I   \\
4414.74 & 7.4     & Fe II & 6F   & 4414.90 & I   \\
4528.23 & 6.6     & Fe II & 6F   & 4528.39 & I   \\
4639.51 & 12.9   & Fe II & 4F   & 4639.68 & I   \\
4664.29 & 4.5     & Fe II & 4F   & 4664.45 & I   \\
4727.91 & 24.9   & Fe II & 4F   & 4728.07 & I   \\
4773.56 & 17.4   & Fe II & 4F   & 4774.74 & I   \\
4814.38 & 54.2   & Fe II & 20F & 4814.49 & I   \\
4874.33 & 16.3   & Fe II & 20F & 4874.49 & I   \\
4889.46 & 33.6   & Fe II & 4F   & 4889.63 & I   \\
4898.44 & 9.7     &         &        &                &     \\
4905.19 & 29.5   & Fe II & 20F & 4905.35 & I   \\
4947.23 & 7.5     & Fe II & 20F & 4947.38 & I   \\
4950.58 & 12.1   & Fe II & 20F & 4950.74 & I   \\
4973.22 & 15.1   & Fe II & 20F & 4973.39 & I   \\
5005.37 & 13.7   & Fe II & 20F & 5005.52 & I   \\
5020.08 & 13.6   & Fe II & 20F & 5020.24 & I   \\
5040.81 & 26.7   &         &        &               &      \\
5043.81 & 6.0     & Fe II & 20F & 5043.53 & I   \\
5107.75 & 7.6     & Fe II & 18F & 5107.95 & III \\
5111.46 & 17.7   & Fe II & 19F & 5111.63 & III \\
5157.83 & 17.9   & Fe II & 18F & 5158.00 & III \\
5158.61 & 53.5   & Fe II & 19F & 5158.81 & III \\
5163.79 & 20.8   & Fe II & 35F & 5163.94 & III \\
5168.85 \footnote{Blended with 5168.73 unidentified} & 9.8     & Fe II & 42   & 5169.03 & III \\
5181.78 & 10.0   & Fe II & 18F & 5181.97 & III \\
5199.00 & 8.0     &         &        &               &     \\
5219.89 & 17.7   & Fe II & 19F & 5220.06 & III \\
5261.46 & 64.4   & Fe II & 19F & 5261.61 & III \\
5268.70 & 15.6   & Fe II & 18F & 5268.88 & III \\
5273.18 & 4.3     & Fe II & 18F & 5273.38 & III \\
5282.94 & 5.9     &         &        &               &     \\
5296.67 & 12.6   & Fe II & 19F & 5296.84 & III \\
5333.48 & 39.7   & Fe II & 19F & 5333.65 & III \\
5347.50 & 4.7     & Fe II & 18F & 5347.67 & III \\
5376.28 & 32.8   & Fe II & 19F & 5376.47 & III \\
5412.49 & 9.1     & Fe II & 17F & 5412.64 & III \\
5414.80 & 4.6     &         &        &               &     \\
5432.99 & 16.6   & Fe II & 18F & 5433.15 & III \\
5477.05 & 9.4     & Fe II & 34F & 5477.25 & III \\
5527.17 & 19.3   & Fe II & 17F & 5527.33 & III \\
5556.18 & 2.1     & Fe II & 18F & 5556.31 & III \\
5580.65 & 4.8     & Fe II & 39F & 5580.82 & III \\
5673.01 & 8.1     & Fe II & F     & 5673.22 & III \\
5746.78 & 22.0   & Fe II & 34F & 5746.96 & III \\
5835.25 & 6.6     & Fe II & F     & 5835.44 & III \\
6364.94 & 9.0     &         &        &               &     \\
6371.07 & 24.1   & Si II  & 2     & 6371.36 & III \\
6666.60 & 13.4   & Ni II  & 2F   & 6668.8   & III \\
6729.64 & 4.0     & Fe II & 31F & 6729.85 & II  \\
6808.99 & 6.0     & Fe II & 31F & 6809.21 & II  \\
6813.41 & 6.7     & Ni II  & 8F   & 6813.73 & II  \\
7154.93 & 51.8   & Fe II & 14F & 7155.16 & IV \\
7171.74 & 18.2   & Fe II & 14F & 7172.00 & IV \\
7377.63 & 73.5   & Ni II  & 2F   & 7377.90 & IV \\
7387.91 & 15.9   & Fe II & 14F & 7388.18 & IV \\
7411.41 & 31.4   & Ni II  & 2F   & 7411.60 & IV \\
7452.30 & 24.5   & Fe II & 14F & 7452.54 & IV \\
7495.36 & 11.3   & Fe II &        & 7495.62 & IV \\
7512.91 & 25.6   & Fe II &        & 7513.16 & IV \\
7731.41 & 8.3     & Fe II &        & 7731.68 & IV \\
7999.80 & 129.4 & Cr II & 1F   & 7999.85 & IV \\
8125.05 & 104.0 & Cr II & 1F   & 8125.22 & IV \\
8184.62 & 2.7     &        &          &               &     \\
8216.04 & 97.8   & N I   & 2      & 8216.28 & IV \\
8308.22 & 60.1   & Cr II & 1F    & 8308.39 & IV \\
8616.65 & 65.1   & Fe II & 13F & 8616.96 & IV \\
8891.62 & 26.0   & Fe II & 13F & 8891.88 & IV \\
8229.43 & 64.6   & Cr II & 1F    & 8229.81 & II  \\
8287.53 & 56.3   &         &         &               &     \\
8694.94 & 31.8   & S I    & 6     & 8694.70 & II   \\
8715.46 & 11.1   & Fe II & 42F & 8715.84 & II   \\
8768.89 & 23.4   &         &        &               &      \\
8819.24 & 17.2   &         &        &               &      \\
9122.64 & 21.9   & Fe II &        & 9122.91 & IV \\
9186.94 & 16.6   & Fe II &        & 9187.15 & IV \\
9203.80 & 42.2   & Fe II &        & 9204.05 & IV \\
9217.98 & 86.7   &         &        &               &     \\
9243.94 & 177.7 &         &        &               &     \\
\end{longtable}

\bsp

\label{lastpage}

\end{document}